# Automated Quantification of CT Patterns Associated with COVID-19 from Chest CT

**Authors:**

Shikha Chaganti, PhD (5), Philippe Grenier, Prof., MD (1), Abishek Balachandran, MD (3), Guillaume Chabin (6), Stuart Cohen, MD (2), Thomas Flohr, apl. Prof., PhD (4), Bogdan Georgescu, PhD (5), Sasa Grbic, PhD (5), Siqi Liu, PhD (5), François Mellot, MD (1), Nicolas Murray, MD (8), Savvas Nicolaou, MD (8), William Parker, MD (8), Thomas Re, MD (5), Pina Sanelli, MD (2), Alexander W. Sauter, MD (7), Zhoubing Xu, PhD (5), Youngjin Yoo, PhD (5), Valentin Ziebandt (4), Dorin Comaniciu, PhD (5)

(1) Hôpital Foch, Suresnes, France
(2) Donald and Barbara Zucker School of Medicine,
    Feinstein Institutes for Medical Research,
    Northwell Health, Manhasset, NY, USA
(3) Siemens Healthinners, Bangalore, India
(4) Siemens Healthineers, Forchheim, Germany
(5) Siemens Healthineers, Princeton, NJ, USA
(6) Siemens Healthineers, Paris, France
(7) University Hospital Basel, Clinic of Radiology & Nuclear medicine, Basel, Switzerland
(8)  Vancouver General Hospital, Vancouver, Canada

## Article Type: Original Research

## Summary

Automated quantification of abnormalities associated with COVID-19 from chest CT could help clinicians evaluate the disease and assess its severity and progression. This study proposes measures of disease severity and a deep learning and deep reinforcement-based method to compute them.

## Key Points

- The method computes percentage of opacity (PO) and lung severity score (LSS) by segmenting three-dimensional abnormalities associated with COVID-19, namely ground glass opacities and consolidations. The PO and LSS quantify the extent of lung involvement and the distribution of involvement across lobes, respectively.
- Given that high opacity abnormalities (ie consolidations and sub-solid regions) were shown to correlate with severe symptoms, this study introduces two complementary severity measures percentage of high opacity (PHO) and lung high opacity score (LHOS). They quantify the extent of high opacity abnormalities and their distribution of involvement across lobes, respectively.
- The combined severity measures of (PO, PHO) and (LSS, LHOS) quantify the extent of overall COVID-19 abnormalities and the presence of high opacities, global and lobe-wise, respectively.

- The performance of the method in estimating PO, LSS, PHO, and LHOS is evaluated on a database of 100 COVID-19 and 100 controls CT scans from multiple institutions from Canada, Europe, and US. Ground truth was established by computing the same measures from manual annotations of the lesions, lungs, and lobes. The Pearson correlation coefficient between method prediction and ground truth for COVID-19 positive scans was calculated as 0.92 for PO ($P < .001$), 0.97 for PHO ($P < .001$), 0.91 for LSS ($P < .001$), 0.9 for LHOS ($P < .001$).

## List of Abbreviations

COVID-19 = Coronavirus Disease 2019, ILD = interstitial lung diseases, LHOS = lung high opacity score, LSS = lung severity score, PHO = percentage of high opacity, PO = percentage of opacity, ROI = region of interest, RT-PCR = reverse transcription polymerase chain reaction, 3D = three-dimensional

# ABSTRACT

## Purpose

To present a method that automatically segments and quantifies abnormal CT patterns commonly present in coronavirus disease 2019 (COVID-19), namely ground glass opacities and consolidations.

## Materials and Methods

In this retrospective study, the proposed method takes as input a non-contrasted chest CT and segments the lesions, lungs, and lobes in three dimensions, based on a dataset of 9749 chest CT volumes. The method outputs two combined measures of the severity of lung and lobe involvement, quantifying both the extent of COVID-19 abnormalities and presence of high opacities, based on deep learning and deep reinforcement learning. The first measure of (PO, PHO) is global, while the second of (LSS, LHOS) is lobe-wise. Evaluation of the algorithm is reported on CTs of 200 participants (100 COVID-19 confirmed patients and 100 healthy controls) from institutions from Canada, Europe and the United States collected between 2002-Present (April, 2020). Ground truth is established by manual annotations of lesions, lungs, and lobes. Correlation and regression analyses were performed to compare the prediction to the ground truth.

## Results

Pearson correlation coefficient between method prediction and ground truth for COVID-19 cases was calculated as 0.92 for PO ($P < .001$), 0.97 for PHO ($P < .001$), 0.91 for LSS ($P < .001$), 0.90 for LHOS ($P < .001$). 98 of 100 healthy controls had a predicted PO of less than 1%, 2 had between 1-2%. Automated processing time to compute the severity scores was 10 seconds per case compared to 30 minutes required for manual annotations.

## Conclusion

A new method segments regions of CT abnormalities associated with COVID-19 and computes (PO, PHO), as well as (LSS, LHOS) severity scores.

# INTRODUCTION

Coronavirus disease (COVID-19) is an infectious disease affecting more than 2 400 000 individuals worldwide as of April 20, 2020 (1). The disease is caused by the severe-acute respiratory syndrome coronavirus 2 (SARS-Cov2) and has a high case fatality rate of up to 4% (2) with more than 170 000 deaths reported currently around the globe. Due to a high reproduction number and infectious nature of the disease, tools for rapid testing and evaluation are vital to track and mitigate the spread.

COVID-19 infection is confirmed by a reverse transcription polymerase chain reaction (RT-PCR) (3). However, the sensitivity of this test may be as low as 60-70% (4) and can result in false negatives within first few days of infection. The role of CT for diagnosis is currently being debated, however two preliminary studies (5,6) showed that chest CT imaging of the lung provides improved sensitivity when compared to RT-PCR for patients suspected to have COVID-19. These studies show that non-contrast chest CT has been useful to detect, quantify severity, and assess progression of the disease. The primary features observed on a lung affected by COVID-19 are peripheral focal or multi-focal ground glass opacities, consolidation, and crazy-paving patterns.

In another study, CT scans were evaluated from patients that tested positive for COVID-19 across 552 hospitals in China (7). They found that 84.4% of patients exhibiting mild symptoms showed CT changes, whereas 94.6% of patients exhibiting severe symptoms showed CT changes. In a study of 121 patients, Bernheim et al showed that CT changes correlated with disease severity, with increasing lung involvement and abnormalities as the disease progressed (8). A recent radiologist performance study also points to preliminary evidence that radiologists can discriminate COVID-19 from other types of pneumonia (9). The study showed that seven United States and Chinese radiologists were able to identify COVID-19 from other types of viral pneumonia with a median sensitivity of 73% and a median specificity of 93%(9). The studies so far have been performed on patients either suspected or confirmed to be tested positive of COVID-19, therefore it is too early to state that CT can be used as an effective diagnostic or screening tool for evaluation of COVID-19 in general population. However, most patients with COVID-19 do seem to exhibit CT abnormalities and the extent of these changes seem to correlate with disease severity. Recently, the Fleischner society provided recommendations for use of chest imaging in evaluation of COVID-19 in regions with resource constraints especially for patients with worsening respiratory status, suspected false negative PCR, and more severe infection. Elsewhere, they indicated that chest imaging can be used to establish a baseline for moderate to severe cases (10). As a result, CT imaging is being adapted as a tool to evaluate COVID-19 in emergency rooms across the world, where other resources for testing are scarce.

We propose an automated system that can quantify the abnormalities most often seen in COVID-19 (ie ground glass opacities and consolidations). The system outputs regions of abnormality demarcated in a three-dimensional CT scan and output two combined severity measures of the disease: *(a)* (percentage of opacity, percentage of high opacity) or (PO, PHO) and *(b)* (lung severity score, lung high opacity score) or (LSS, LHOS). The first measure (PO,PHO) indicates the overall disease spread relative to the volume of the lungs, while the second measure (LSS, LHOS) is a cumulative measure of the extent of involvement for each lung lobe (Figure 1). Opacities here include GGO, consolidation, and crazy-paving patterns which are all commonly observed in COVID-19. Given that higher opacity abnormalities such as consolidation were shown to correlate with severe symptoms, the measures cover both the extent of disease and the presence of high opacities (7,11). In this study, we use the term *high opacity* to mean consolidation of sub-solid regions of high attenuation. We use this term since these regions were determined using an empirical threshold. The ability to determine the extent of abnormalities and quantify the presence of high opacities provides valuable insight for prognosis prediction, risk prioritization, and therapy response (12).

## MATERIALS AND METHODS

All the authors have either been employed or partially supported by Siemens Healthineers.

### Datasets and Annotation.

The datasets were de-identified and retrospectively collected from multiple centers in USA, Canada, and Europe, between 2002 to present (April, 2020) with the data usage being approved by the respective ethics committees, with waived consent. A subset of the datasets are from NLST (13) and COPDgene (14). Non-contrasted chest CT datasets that completely show lungs in their field of view were used to train two models: one for lung and lobe segmentation and another for abnormality segmentation. Complete details of the datasets are provided in Tables 1 and 2, including age and sex distribution, CT scanner manufacturer, slice thickness, and kernel type.

A total of 9749 CT volumes were used in the study. A total of 200 volumes were reserved for testing, and the remaining were used for lung and lesion segmentation training and model selection. A total of 901 volumes were used for training the abnormality segmentation model. Due to limited availability of COVID-19 data, we determined three groups of data that are useful for training the abnormality model: COVID-19 datasets ($n$ = 431), pneumonia datasets ($n$ = 174) and other interstitial lung diseases (ILDs, $n$ = 274). Note that 575 datasets were common to the lung segmentation and lesion segmentation training sets. The distribution of the datasets for each task is visualized in Figure 2. Annotation details for the two tasks are presented in Appendix E1 (supplement). A separate test set of 100 control volumes and 100 COVID-19 positive volumes were used to evaluate lung and lobe segmentation as well as abnormality segmentation. The control group consisted of volumes without CT abnormalities that were randomly sampled and visually inspected by a radiologist. There were 110 COVID-19 positive cases that were randomly selected from two European and two institutions from the United States. Of these, 10 cases were excluded for having very low percentage of abnormalities based on ground truth annotations.

## Lung and Lobe Segmentation

The algorithm generates lung and lobe segmentation mask on a given CT data set. First, multi-scale deep reinforcement learning (15) is used to robustly detect anatomical landmarks in a CT volume. The carina bifurcation is used to identify the lung region of interest (ROI). When the lung ROI is not detected by the system, the sternum tip is used as a landmark. The size and the relative location of the lung ROI to the landmark are specified according to annotated lung datasets. Next, the lung ROI image is resampled to a 2 mm isotropic volume and fed into a Deep Image-to-Image Network (DI2IN) (16) to generate the lung segmentation. Finally, the ROI segmentation mask is remapped to have the same dimension and the resolution as the input data. The DI2IN has been first trained on CT scans from a large group of patients with various diseases (miscellaneous in Table 1), then fine-tuned with scans with abnormal patterns including ( ILDs, pneumonia, and COVID-19 in Table1), to improve the robustness of the lung segmentation over the infected areas.

## COVID-19 Related Abnormality Segmentation

We formulated the segmentation of the COVID-19 related abnormality patterns as a semantic segmentation problem involving binary classes. A DenseUNet (17) with anisotropic kernels is trained to transfer a three-dimensional (3D) chest CT volume to a semantic segmentation mask of the same size. We use a single label to define all voxels within the lungs that fully or partially contain ground glass opacity or consolidations as the positive voxels (see Annotation Details in Appendix E1). The rest of the image areas within the lungs and the entire area outside the lungs are defined as negative. The network is trained as an end-to-end segmentation system. The 3D output segmentation is filtered by the lung segmentation as described above. The technical details of the network architecture and the training strategy are summarized in Appendix E1 (supplement).

## COVID-19 Severity Measures

The severity of airspace abnormalities in each subject is measured by the combined metrics (PO, PHO) and (LSS, LHOS) as defined below.

Percentage of Opacity (PO): The PO is calculated as the total percent volume of the lung parenchyma that is affected by disease:

$$PO = 100 \times \frac{volume\ of\ predicted\ abnormalities}{volume\ of\ lung\ mask}$$

Percentage of High Opacity (PHO): The PHO is calculated as the total percent volume of the lung parenchyma that is affected by severe disease (ie high opacity regions including consolidation). The high-opacity regions were identified using a threshold on -200 HU, from the abnormality segmentation:

$$PHO = 100 \times \frac{volume\ of\ high\ opacity\ region}{volume\ of\ lung\ mask}$$

Lung Severity Score (LSS): The LSS is computed to measure the extent of lung involvement across each lobe as described by Bernheim et al (8). For each lobe, the percentage of affected lobe is calculated and scored between 0-4: (0) lobe is not affected; (1) 1-25% of the lobe is affected; (2) 25-50% of the lobe is affected; (3) 50-75% of the lobe is affected; and (4) 75-100% of the lobe is affected.

The scores for each of the five lobes are summed to calculate the total LSS, resulting in a total score range from 0-20. A 0 indicates that none of the lobes are involved and 20 indicates that all five lobes are severely affected.

Lung High Opacity Score (LHOS): The LHOS is computed to measure the extent of abnormalities with severe disease. For each lobe, the percentage of high opacity is calculated and scored between 0-4: (0) lobe is not affected, or abnormalities do not have any consolidation; (1) 1-25% of the lobe has high opacity abnormalities; (2) 25-50% of the lobe has high opacity abnormalities; (3) 50-75% of the lobe has high opacity abnormalities; and (4) 75-100% of the lobe has high opacity abnormalities.

The scores for each of the five lobes are summed to calculate the total LHOS, resulting in a total score range from 0-20. A 0 indicates that none of the lobes have a high opacity abnormality and 20 indicates that all five lobes are severely affected by high opacity consolidations.

## Statistical Analysis

The aim of the study is to show that abnormalities associated with COVID-19, namely GGO and consolidation, are segmented in a given CT scan, and to also demonstrate that there are no false positive segmentations when there are no abnormalities. In order to do this, we show that ground truth matches the predicted scores of severity using the following statistical tests.

### Chi-squared contingency test

We compute the chi-squared contingency test (18) between ground truth and predicted scores. For LSS and LHOS, the frequencies are computed for each value from 0-20 in order to calculate the test. For PO and PHO the frequencies are estimated based on clinically relevant (8) bins, 0-1%, 1-25%, 25-50%, 50-75%, 75-100%.

### Correlation

We compute two types of correlations: Pearson's coefficient and Kendall's Tau using python's scipy package (19,20). Pearson's coefficient demonstrates the linear association between ground truth and predicted values. The Kendall's Tau measures the ordinal relationship between the ground truth and predicted metrics, i.e., it measures the agreement between the rankings of the two measures. We compute the correlations for COVID-19 positive subjects.

### Linear relationship

In order to further evaluate the linear relationship between the ground truth and predicted metrics, we compute the linear regression estimates for metrics which have continuous values (PO and PHO), as follows:

$$metric_{gt} = \beta_0 + \beta_1 metric_{predicted}$$

We report the following:

- $\hat{\beta}_0$, the estimate of the intercept of the linear regression, along with the confidence interval.
- $\hat{\beta}_1$, the estimate of the slope for the linear regression, along with the confidence interval.
- Coefficient of determination or R-squared value.
- Mean squared error of the metric from the regression line.

# RESULTS

## Training of Lung and Lobe Segmentation

We used 8792 chest CT volumes for training the 3D segmentation of lung and lobes. The DI2IN was first trained on a data set from 8087 patients (8006 for training and 81 for validation) without the prevalence of viral pneumonia, then it was fine-tuned on another data set from 1136 patients (1076 for training and 60 for validation) with abnormality patterns including ground glass opacities, consolidation, effusions, masses, and others to improve the robustness of the lung and lobe segmentation over the infected areas. The DI2IN was finally tested on a data set from 200 patients, consisting of 100 COVID-19 and 100 controls (13). The sample size and the data partition of the COVID-19 positive patients are determined by balancing the patients from the US and Europe (see Table 1).

## Training of Abnormality Segmentation

For the training of abnormality segmentation, we used a combination of COVID19, viral pneumonia, and other ILD datasets, totaling 901 CT scans (431 COVID-19, 174 viral pneumonia, and 296 ILD). Atypical pneumonia datasets including SARS, MERS, and other viral pneumonia have similar CT abnormalities as COVID-19 (21) and are used as reasonable proxies to complement the COVID-19 training dataset together with CT datasets that include other ILDs that present with GGO and consolidation. These datasets were used in the training dataset to learn the patterns relevant in COVID-19. For the testing set, both the initial annotation and final review was done by a board-certified radiologist. For testing, we used the same 200 datasets as for the lung and lobe segmentation, which includes 100 COVID-19 and 100 controls (13).

## Computation of Combined Measures of Disease Severity

From the predicted abnormality segmentation, we compute the two proposed combined measures of disease severity: *(a)* (PO, PHO) and *(b)* (LSS, LHOS) for the 200-test data set that consisted of 100 COVID-19 and 100 controls. For COVID-19 cases we show that the abnormalities of GGO and consolidation are segmented and quantified with a high correlation to the ground truth annotations. We also show that there are no false positives segmented in most subjects with no pathology. A chi-squared contingency test between ground truth and predicted metrics shows that there is no significant difference between ground truth and predicted metrics for clinically relevant binning of the measures (see Table 3). 98 of the 100 healthy controls have a predicted PO less than 1%. 2 of them have PO detected between 1-2%. We also computed linear and ordinal correlations for the four metrics. The Pearson coefficients for COVID-19 positive cases are 0.92 for PO ($P < .001$); 0.97 for PHO ($P < .001$); 0.91 for LSS ($P < .001$); and 0.90 for LHOS ($P < .001$). The Kendall Tau is 0.77 for PO ($P < .001$); 0.86 for PHO ($P < .001$); 0.83 for LSS ($P < .001$); and 0.83 for LHOS ($P < .001$). All correlations are statistically significant. Figure 4 shows the scatter plot between ground truth and predicted metrics PO, PHO, LSS, and LHOS. Figure 5 shows the combined measures of (PO, PHO) as well as (LSS, LHOS) in two-dimensional plots. The diagonal in Figure 5(a) indicates that PO is equal to PHO. A subject's marker falling on the diagonal would indicate that 100% of the abnormalities present on their lung CT are high opacity or consolidation. Note that all the markers for abnormalities greater than 0 fall below the line, indicating various levels of consolidation present. The closer a subject's marker is to the diagonal, the more consolidation there is in their lungs. In Figure 5(b), the diagonal indicates that LSS is equal to LHOS. A marker on the diagonal here indicates that all the lobes that are involved have high opacity. When a patient's marker falls below the diagonal, it means that either some lobes do not have high opacities or that they have less extent of high opacities.

In addition to correlation, we also present estimates of the linearity of the relationship between ground truth and predicted metrics for PO and PHO for COVID-19 positive cases, as shown in Table 4. The intercepts are 0.03 and 0.004, and the slopes are 0.839 and 0.805 for PO and PHO respectively. The mean error between ground truth and predicted PO and PHO are 3.5% and 1.9% respectively. Figure 6 shows the qualitative results for COVID-19 positive cases along with their scores. Figure 6 (a) shows six cases where the automated segmentation of the algorithm matches the ground truth closely. For example, the upper left corner case resulted in a (PO, PHO) = (55.9%, 7.6%). That means that 55.9% of the lung volume is affected by disease and 7.6% of the lung volume has high consolidation abnormalities. Figure 6 (b) shows one example of abnormality segmentation (left) along with high opacity regions identified through thresholding (right). Figure 6 (c) shows two cases which were identified as outliers due to their PO score differing by more than 20% from the ground truth. The algorithm misses fainter areas of opacity (left) and tends to over segment into normal areas between the abnormal opacities (right).

# DISCUSSION

In this work we proposed a method to automatically quantify regions with ground glass opacity and consolidation in chest CT scans using deep learning algorithms. The abnormalities segmented by our system are the most common chest CT findings in COVID-19. The system outputs the 3D contours of the CT abnormalities and computes two combined severity measures of (PO, PHO) and (LSS, LHOS).

## Clinical value of our system

Automated characterization of disease can be extremely useful to facilitate rapid response in assessing the extent of lung infection and progression of the disease in patients exhibiting symptoms of COVID-19. Evaluation of a chest CT can be used to answer several questions in a clinical setting such as triage of a patient, diagnosis, assessment of severity and progression, and response to therapy. Recently, lung severity measures have been proposed for COVID-19 that demonstrate correlation to disease severity and progression (8,22–25). Manual calculation of these scores produces a two-fold problem: either the affected regions are annotated precisely, which is a time-consuming process, or the affected regions are assessed subjectively, which translates into low reproducibility of the assessment. A precise automated measurement of lung severity scores will address both the speed and reproducibility problems. In addition to quantitative outputs, the abnormality segmentation produced by the system can also be used for visual interpretation and inspection.

The automatic computation of (PO, PHO) as well as (LSS, LHOS) as provided by our system would be useful in several clinical scenarios:

Severity assessment.— An objective quantitative measure of severity is provided by (PO, PHO) and (LSS, LHOS) scores. This measure can be used in triage, to prioritize patients requiring hospital or intensive care unit admittance, and to assess the prognosis of a patient.

Progression assessment.— In several instances, the patient might be imaged multiple times to assess disease progression or to assess response to drugs. In order to make these decisions, an objective metric or a score that can assess the extent of disease is vital. The (PO, PHO) and (LSS, LHOS) scores can be useful when evaluating a patient's lung status over time. Such a tool could also be used in scientific research studies that are documenting the temporal radiological change of COVID-19 pneumonia and clinical trials studying drug responses.

Early detection.— Although the evidence is preliminary, the system could be used as an effective tool in conjunction with RT-PCR tests to increase overall sensitivity of COVID-19 detection, especially in scenarios where there are limited resources and a crunch for time.

An automated system that can perform this disease severity assessment can be of great use to hospitals that have a shortage of manpower combined with a high-volume of patients during the outbreak (12).

## Technical Contributions

To our knowledge this is one of the first systems to provide an automatic, efficient, and detailed evaluation of tomographic CT patterns associated with COVID-19. We also believe that this is the first system to propose and show how to automatically compute severity measures such as (PO, PHO) and (LSS, LHOS) that characterize not only the extent of disease in the lungs, but also the extent of high opacity abnormalities, both globally and lobe-wise. This performance is facilitated by the experience of a multidisciplinary team of clinicians, machine learning experts, and engineers, whose combined clinical and technical knowledge helped to solve robust landmark detection in chest CT, 3D lung and lobe segmentation, and 3D segmentation of abnormalities associated with COVID-19.

## Limitations and Future Work

The system was trained with specific abnormalities that were relevant to COVID-19 such as ground glass capacities and consolidations. Presence of other abnormalities in or around the lung such as pleural effusion could produce challenges to the algorithm. In the future, the system will be trained comprehensively to account for abnormalities other than GGO and consolidation. The algorithm will be improved using more training data in boundary conditions, which cause under segmentation with faint regions and over segmentation with highly abnormal regions as seen in Figure 6 (c). The slope and intercept in table 4 show that the AI system outputs a value that is slightly higher than ground truth value, indicating that the system needs to be calibrated further for accurate metric measurements. The system produces a quantification of abnormalities in a lung CT scan only. While we evaluate this quantification in COVID-19 and healthy cases, we do not evaluate it in other lung diseases such as other viral pneumonias or ILDs. It is difficult to ascertain the differences between diseases and diagnosis capability based on this information alone. In the future, we will evaluate the feasibility of distinguishing COVID-19 from other types of pneumonias and ILDs. We will conduct studies to correlate the imaging findings computed in this work to clinical factors and outcomes. A similar study can be performed with X-ray chest images as well, which will be a part of our future investigations. These additional studies can provide a comprehensive toolbox for clinicians to diagnose and treat COVID-19.

Disclaimer: The concepts and information presented in this paper are based on research results that are not commercially available

# ACKNOWLEDGEMENTS

We gratefully acknowledge the contributions of Clinica Universidad de Navarra, Health Time, Houston Methodist, and multiple other frontline hospitals to this collaboration. The authors also thank the COPDGene for providing the data. The COPDGene study (NCT00608764) was funded by NHLBI U01 HL089897 and U01 HL089856 and also supported by the COPD Foundation through contributions made to an Industry Advisory Committee comprised of AstraZeneca, BoehringerIngelheim, GlaxoSmithKline, Novartis, and Sunovion.  We thank many colleagues who made this work possible in a short amount of time. Special recognition to Christian Eusemann, and Ahmed Halaweish, for the great effort to ensure proper data de-identification and compliance with local regulations, to Dr. Vishwanath RS and Dr. Eileen Krieg for their valuable input on CT data and their annotations, to Brian Teixeira and Sebastien Piat, who were instrumental for implementing and managing the data infrastructure, and to Marlène Roblot, for her support with the manuscript editing.

# REFERENCES

1. JHU. Coronavirus COVID-19 Global Cases by the Center for Systems Science and Engineering (CSSE) at Johns Hopkins University (JHU) [Internet]. 2020. Available from: https://gisanddata.maps.arcgis.com/apps/opsdashboard/index.html#/bda7594740fd40299423467b48e9ecf6

2. Wilson N, Kvalsvig A, Barnard LT, Baker MG. Case-Fatality Risk Estimates for COVID-19 Calculated by Using a Lag Time for Fatality. Emerg Infect Dis. 2020;26(6).

3. CDC. CDC Clinical Criteria [Internet]. 2020. Available from: https://www.cdc.gov/coronavirus/2019-nCoV/hcp/clinical-criteria.html

4. Kanne JP, Little BP, Chung JH, Elicker BM, Ketai LH. Essentials for radiologists on COVID-19: an update—radiology scientific expert panel. Radiology. 2020;200527.

5. Fang Y, Zhang H, Xie J, Lin M, Ying L, Pang P, et al. Sensitivity of chest CT for COVID-19: comparison to RT-PCR. Radiology. 2020;200432.

6. Xie X, Zhong Z, Zhao W, Zheng C, Wang F, Liu J. Chest CT for Typical 2019-nCoV Pneumonia: Relationship to Negative RT-PCR Testing. Radiology [Internet]. 2020;200343. Available from: http://www.ncbi.nlm.nih.gov/pubmed/32049601

7. Guan W, Ni Z, Hu Y, Liang W, Ou C, He J, et al. Clinical characteristics of coronavirus disease 2019 in China. N Engl J Med. 2020;

8. Bernheim A, Mei X, Huang M, Yang Y, Fayad ZA, Zhang N, et al. Chest CT findings in coronavirus disease-19 (COVID-19): relationship to duration of infection. Radiology. 2020;200463.

9. Bai HX, Hsieh B, Xiong Z, Halsey K, Choi JW, Tran TML, et al. Performance of radiologists in differentiating COVID-19 from viral pneumonia on chest CT. Radiology. 2020;200823.

10. Rubin GD, Ryerson CJ, Haramati LB, Sverzellati N, Kanne JP, Raoof S, et al. The Role of Chest Imaging in Patient Management during the COVID-19 Pandemic: A Multinational Consensus Statement from the Fleischner Society. Chest. 2020;

11. Inui S, Fujikawa A, Jitsu M, Kunishima N, Watanabe S, Suzuki Y, et al. Chest CT findings in cases from the cruise ship "Diamond Princess" with coronavirus disease 2019 (COVID-19). Radiol Cardiothorac Imaging. 2020;2(2):e200110.

12. Kim H. Outbreak of novel coronavirus (COVID-19): What is the role of radiologists? Springer; 2020.

13. Team NLSTR. The national lung screening trial: overview and study design. Radiology. 2011;258(1):243–53.

14. Arroz SDE, Usos CON, En P, Salud AY. Study Protocol: COPDGene. 2009;(1):2.

15. Ghesu F-C, Georgescu B, Zheng Y, Grbic S, Maier A, Hornegger J, et al. Multi-scale deep reinforcement learning for real-time 3D-landmark detection in CT scans. IEEE Trans Pattern Anal Mach Intell. 2017;41(1):176–89.

16. Yang D, Xu D, Zhou SK, Georgescu B, Chen M, Grbic S, et al. Automatic liver segmentation using an adversarial image-to-image network. Lect Notes Comput Sci (including Subser Lect Notes Artif

Intell Lect Notes Bioinformatics). 2017;10435 LNCS:507–15.

17. Ronneberger O, Fischer P, Brox T. U-net: Convolutional networks for biomedical image segmentation. In: International Conference on Medical image computing and computer-assisted intervention. 2015. p. 234–41.

18. Cressie N, Read TRC. Multinomial goodness-of-fit tests. J R Stat Soc Ser B. 1984;46(3):440–64.

19. Virtanen P, Gommers R, Oliphant TE, Haberland M, Reddy T, Cournapeau D, et al. SciPy 1.0: fundamental algorithms for scientific computing in Python. Nat Methods. 2020;17(3):261–72.

20. Knight WR. A computer method for calculating Kendall's tau with ungrouped data. J Am Stat Assoc. 1966;61(314):436–9.

21. Hosseiny M, Kooraki S, Gholamrezanezhad A, Reddy S, Myers L. Radiology Perspective of Coronavirus Disease 2019 (COVID-19): Lessons From Severe Acute Respiratory Syndrome and Middle East Respiratory Syndrome. Am J Roentgenol. 2020;1–5.

22. Chung M, Bernheim A, Mei X, Zhang N, Huang M, Zeng X, et al. CT Imaging Features of 2019 Novel Coronavirus (2019-nCoV). Radiology. 2020;200230.

23. Pan F, Ye T, Sun P, Gui S, Liang B, Li L, et al. Time Course of Lung Changes On Chest CT During Recovery From 2019 Novel Coronavirus (COVID-19) Pneumonia. Radiology. 2020;200370.

24. Zhao W, Zhong Z, Xie X, Yu Q, Liu J. Relation Between Chest CT Findings and Clinical Conditions of Coronavirus Disease (COVID-19) Pneumonia: A Multicenter Study. AJR Am J Roentgenol [Internet]. 2020;(October):1–6. Available from: http://www.ncbi.nlm.nih.gov/pubmed/32125873

25. Yang R, Li X, Liu H, Zhen Y, Zhang X, Xiong Q, et al. Chest CT Severity Score: An Imaging Tool for Assessing Severe COVID-19. Radiol Cardiothorac Imaging. 2020;2(2):e200047.

26. Huang G, Liu Z, van der Maaten L, Weinberger KQ. Densely Connected Convolutional Networks. arXiv e-prints. 2016 Aug;arXiv:1608.06993.

27. Luo L, Xiong Y, Liu Y, Sun X. Adaptive Gradient Methods with Dynamic Bound of Learning Rate. arXiv e-prints. 2019 Feb;arXiv:1902.09843.

# TABLES

**Table 1.** Properties of Training and Testing Data used for Lung and Lobe Segmentation

| Parameter | Training dataset | Testing Dataset |
|---|---|---|
| Dataset composition | | |
|   Total | 9223 | 200 |
|   COVID-19 | 148 | 100 |
|   Viral Pneumonia | 161 | 0 |
|   Interstitial Lung Disease | 827 | 0 |
|   Miscellaneous | 8087 | 0 |
|   Normal | | 100 |
| Sex | | |
|   Women | 3966 | 71 |
|   Men | 4570 | 112 |
|   Unknown | 687 | 17 |
| Age (y) | | |
|   Median (interquartile range) | 57 (45-66) | 61 (56-63) |
|   Unknown age (*n*) | 5126 | 64 |
| Manufacturer | | |
|   Siemens | 4719 | 62 |
|   GE | 3406 | 68 |
|   Philips | 528 | 24 |
|   Toshiba | 2 | 28 |
|   Unknown or other | 568 | 18 |
| Slice thickness (mm) | | |
|   ≤ 1.5 | 8674 | 56 |
|   1.5-3.0 | 395 | 124 |
|   > 3.0 | 154 | 20 |
| Reconstruction kernel | | |
|   Soft | 8393 | 119 |
|   Hard | 179 | 81 |
|   Unknown | 651 | 0 |

Note.— Both data from the training and testing sets were derived from multiple clinical sites in Europe, Canada, and the United States.

**Table 2.** Properties of Training and Testing Data used for the Segmentation of Abnormalities Associated with COVID-19

| Parameter | Training dataset | Testing dataset |
| --- | --- | --- |
| Dataset composition | | |
|   Total | 901 | 200 |
|   COVID-19 | 431 | 100 |
|   Viral Pneumonia | 174 | 0 |
|   Interstitial Lung Disease | 296 | 0 |
|   Normal | 0 | 100 |
| Sex | | |
|   Women | 247 | 71 |
|   Men | 300 | 112 |
|   Unknown | 354 | 17 |
| Age (y) | | |
|   Median (interquartile range) | 60 (51-68) | 61 (56-63) |
|   Unknown | 533 | 64 |
| Manufacturer | | |
|   Siemens | 485 | 62 |
|   GE | 201 | 68 |
|   Philips | 34 | 24 |
|   Toshiba | 17 | 28 |
|   Unknown or other | 164 | 18 |
| Slice thickness (mm) | | |
|   ≤ 1.5 | 533 | 56 |
|   1.5-3.0 | 285 | 124 |
|   > 3.0 | 83 | 20 |
| Reconstruction kernel | | |
|   Soft | 111 | 119 |
|   Hard | 489 | 81 |
|   Unknown | 301 | 0 |

Note.— Both data from the training and testing sets were derived from multiple clinical sites in Europe, Canada, and the United States.

**Table 3.** Pearson's Correlation Coefficient Between Predicted Disease Severity Measures and Measures Derived from Ground Truth for 100 COVID-19 Positive and 100 Control Test Cases

|  | Chi-squared contingency test (COVID-19 and Control) | | Pearson's Correlation (COVID-19 only) | | Kendall's Rank Correlation (COVID-19 only) | |
| --- | --- | --- | --- | --- | --- | --- |
| Metric | Chi2 | P value | Coefficient | P value | Tau | P value |
| PO | 2.11 | 0.72 | 0.92 | <.001 | 0.77 | <.001 |
| PHO | 1.00 | 0.60 | 0.97 | <.001 | 0.86 | <.001 |
| LSS | 23.50 | 0.2 | 0.91 | <.001 | 0.83 | <.001 |
| LHOS | 2.21 | 0.95 | 0.9 | <.001 | 0.83 | <.001 |

Note.— COVID-19 = coronavirus disease 2019, LHOS = lung high opacity score, LSS = lung severity score, PHO = percentage of high opacity, PO = percentage of opacity

**Table 4.** Model Fitting Parameters for Predicted Disease Severity Measures and Measures Derived from Ground Truth for 100 COVID-19 Positive only

|  | Model Fit | | | |
| --- | --- | --- | --- | --- |
| Metric | R2 | Mean error | Intercept (estimate, confidence interval) | Slope (estimate, confidence interval) |
| PO | 0.854 | 3.5% | 0.033, [0.013, 0.054] | 0.839, [0.769, 0.908] |
| PHO | 0.948 | 1.9% | 0.004, [0.002, 0.007] | 0.805, [0.768, 0.843] |

Note.— PHO = percentage of high opacity, PO = percentage of opacity

# FIGURES

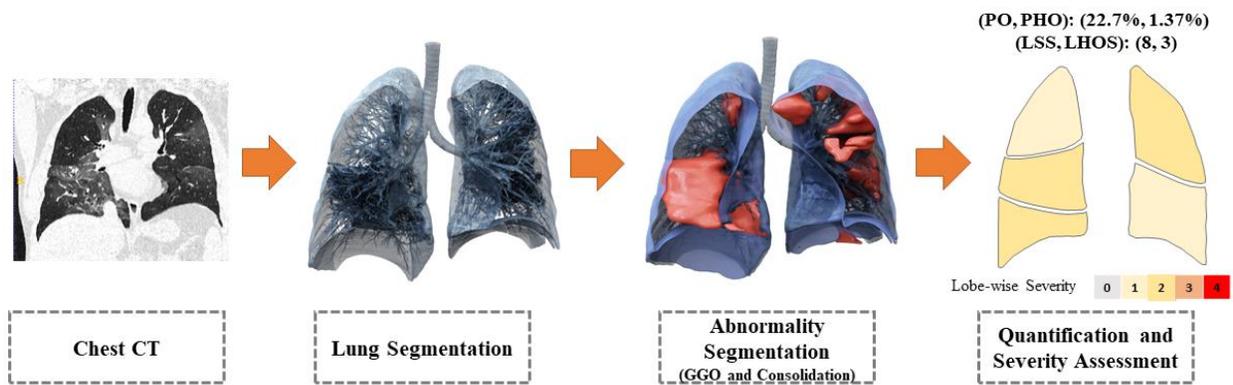

Figure 1. The system takes as input a non-contrasted chest CT and segments the lesions, lungs, and lobes into three dimensions. It outputs two combined measures of the severity of lung and lobe involvement, quantifying both the extent of COVID-19 abnormalities and presence of high opacities. The first measure (Percentage of Opacity, Percentage of High Opacity) or (PO, PHO) is global, while the second measure consists of (Lung Severity Score, Lung High Opacity Score) or (LSS, LHOS) is lobe-wise (see section COVID-19 Severity Metrics).  GGO = ground-glass opacities

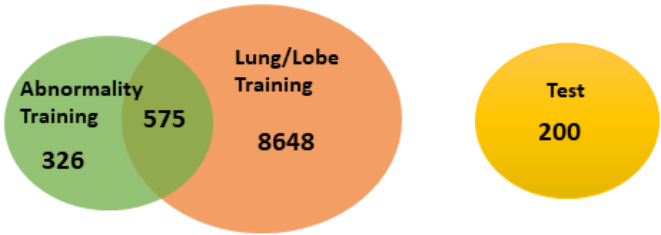
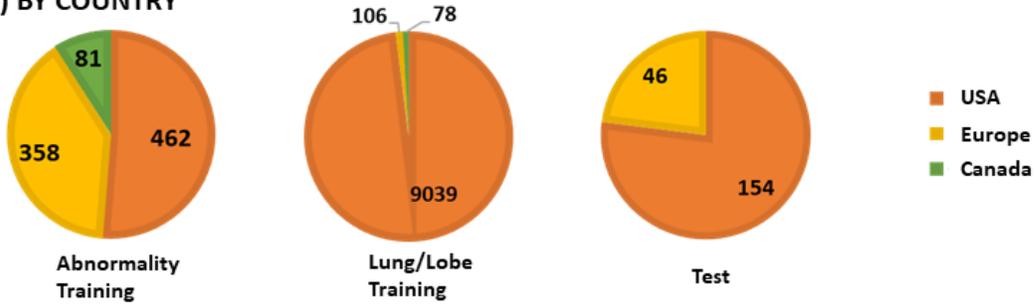
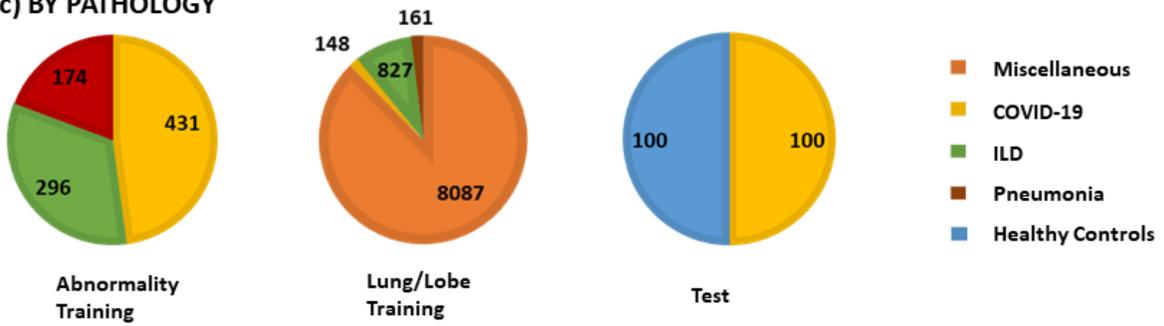

Figure 2. The training and testing set composition in terms of country of origin and lung pathology. ILD = interstitial lung disease

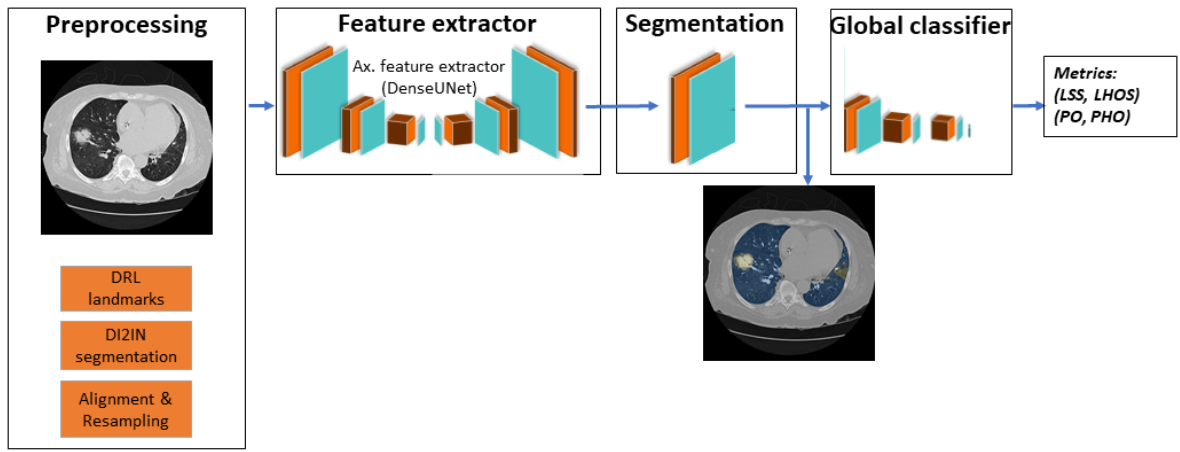

Figure 3. Overview of the deep learning system implemented in this work. LHOS = lung high opacity score, LSS = lung severity score, PHO = percentage of high opacity, PO = percentage of opacity

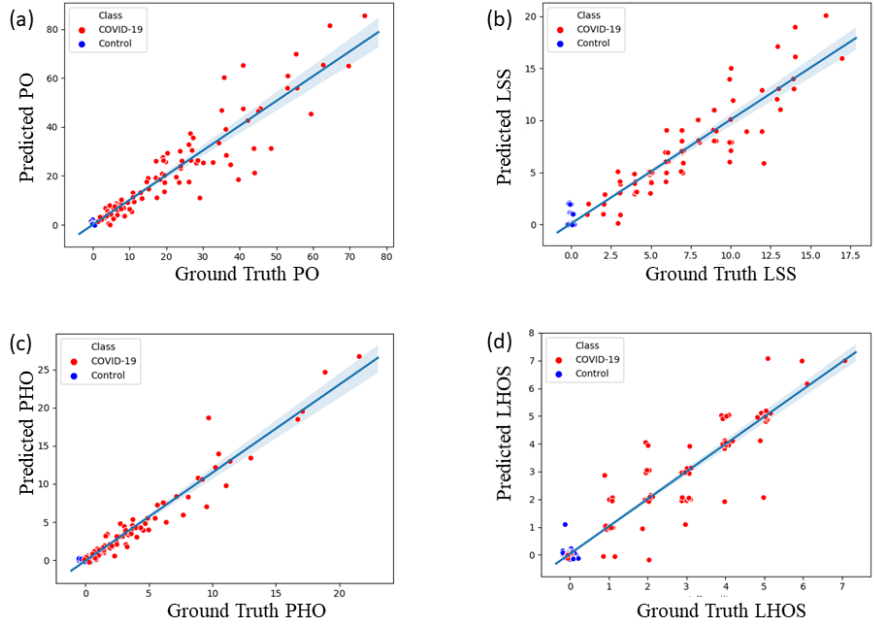

Figure 4. Scatter plots for the four metrics computed on 100 COVID-19 cases and 100 controls. **(a)** Ground truth vs. predicted percentage of opacity (PO). **(b)** Ground truth vs. predicted lung severity score (LSS). **(c)** Ground truth vs. predicted percentage of high opacity (PHO). **(d)** Ground Truth vs. predicted lung high opacity score (LHOS). Note that a small jitter (0.2%) is added for the purpose of visualization to represent overlapping points.

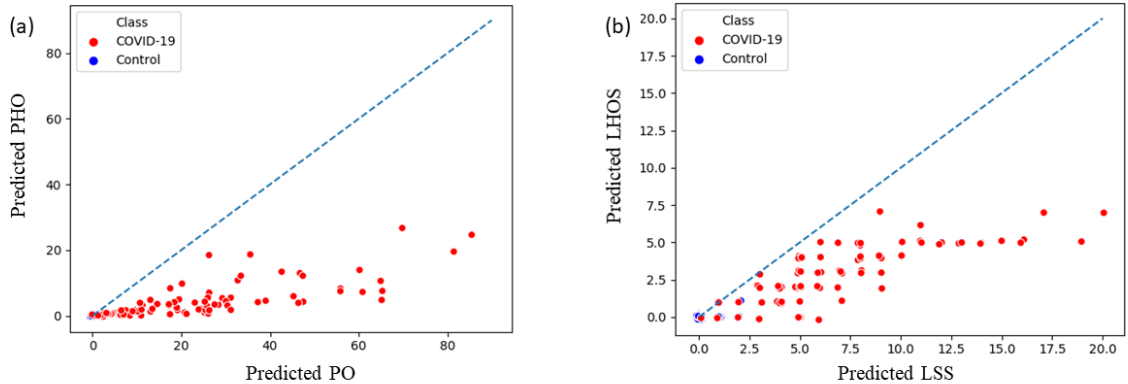

Figure 5. Combined severity measures of **(a)** (Percentage of Opacity, Percentage of High Opacity) or (PO, PHO ) and **(b)** (Lung Severity Score, Lung High Opacity Score) or (LSS, LHOS) computed on 100 COVID-19 cases and 100 normal control CT scans. PO indicates the extent of opacity and PHO indicates the extent of the abnormality that has a high opacity, corresponding to consolidation regions. **(a)** The diagonal line indicates PO and PHO are equal (ie the abnormalities have 100% consolidation). The closer a marker is to the diagonal, the higher the relative percentage of consolidation in that . **(b)** The diagonal indicates that all the lobes that are affected have a high percentage of consolidation as well. The points below the diagonal indicate that some of the lobes affected do not have as severe consolidation as the others. (see section COVID-19 Severity Metrics). Note that a small jitter (0.2%) is added for the purpose of visualization to represent overlapping points.

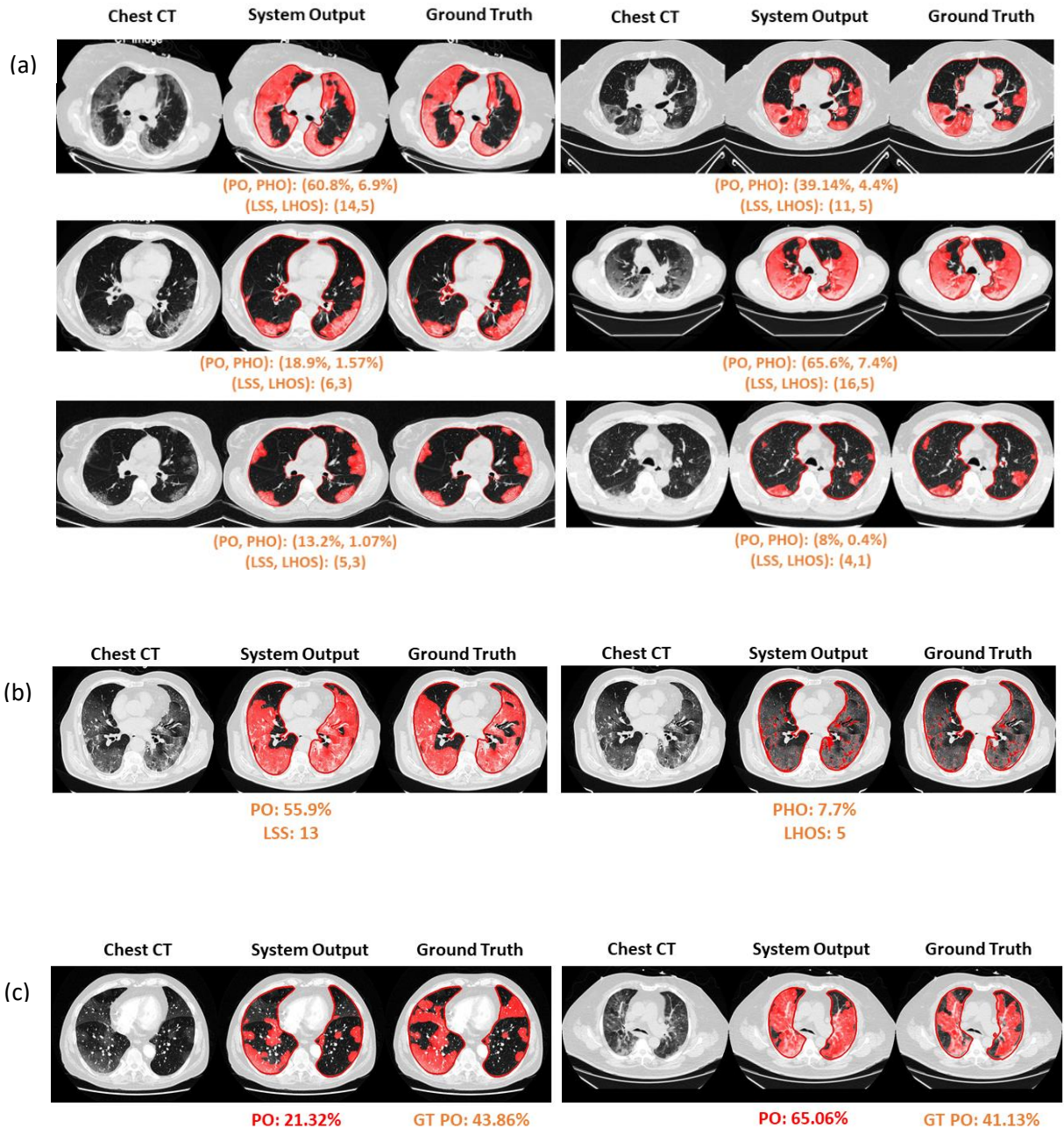

Figure 6. Visualization of segmented abnormality regions. **(a)** Combined severity measures (PO, PHO) and (LSS, LHOS) computed for six successful cases. **(b)** Shows the same result side by side for abnormality segmentation, and just the high opacity segmentation. **(c)** Shows some of the outliers of the system.

# APPENDIX E1

## Annotation Details

ITK-Snap was used for ground truth annotation of lung, lobe, and abnormality segmentation task. The window level and width are adjusted to 425 and 1500. The annotation task was performed by a team of 25 expert medical annotation engineers with a median of 2.25 years of experience of annotating medical imaging studies (range: 1-3.5 years). The tasks were performed under the supervision of 3 board-certified radiologists with 2, 5, and 24 years of experience. Every dataset was initially annotated by our annotators and later reviewed and corrected by a radiologist for any areas missed or to be removed.

The lung lobe masks were initialized by a pre-trained segmentation model. The mask follows the lung boundary against the mediastinum and soft tissue chest wall. The lung lobe segmentation mask covers the lung parenchyma from the division of the primary to secondary bronchus onwards. The mask adjusted to lung parenchymal window for best visualization of opacities includes any areas of consolidation, collapse and other abnormalities. The lung lobe annotation was performed by separating the whole lung into three right lobes and two left lobes at the fissure boundaries with different mask labels.  A subset of the training dataset of 8087 chest CT was annotated previously as a part of the COPDgene study (14).

For the ground truth of the abnormality, a single label mask included all areas of ground glass opacity and consolidation within both the lung fields. However, pleural effusion was excluded in the mask. For crazy paving pattern the mask excluded normal areas of the lung with precision contouring of the mask.

The average time for annotation and review of the lung and lobe masks is 2 hours.

## Abnormality Segmentation Network Architecture

First, the training images are resampled to the resolution $1 \times 1 \times 3mm$. The image intensity is clipped using the standard lung window with the width 1500 HU and level −600 HU before being normalized to [0,1]. We use the predicted lung masks described in Sec. II-B to compute the geometric center of the lungs and then crop the image with a fixed bounding box of size $384 \times 384 \times 384$. For data augmentation, we perturb the image intensity within a random interval [−20,20] and then flip the image in one of the three dimensions with equal probability. The tensor 3D dimensions are kept in z-y-x order throughout training and inference.

We use a UNet (17) resembling architecture with convolutional blocks containing either $1 \times 3 \times 3$ or $3 \times 3 \times 3$ CNN kernels to deal with the anisotropic image resolutions. The network architecture is illustrated in Figure 3. The 3D input tensor is first fed into a 3D $1 \times 3 \times 3$ convolutional layer followed by batch normalization and LeakyReLU. The features are then propagated to 5 DenseNet blocks (26). For the first two DenseNet blocks, the features are downsampled by a $1 \times 2 \times 2$ convolution with a stride of $1 \times 2 \times 2$. These anisotropic downsampling kernels are designed to preserve the inter-slice resolution of the input tensors. The last 3 DenseNet blocks have isotropic downsampling kernels with the stride of $2 \times 2 \times 2$. The input to each decoder block is obtained by concatenating the encoder output features with the same resolution and the features upsampled from the previous decoder. The upsampling kernels are built with transpose convolutional kernels with the sizes and strides same to the corresponding DenseNet blocks. The final network output is derived by projecting the feature maps to 2 feature channels and being activated by the softmax activation. The network parameters are randomly initialized.

We use the Jaccard index as the training loss function. The loss function $L(p, y)$ between the probability prediction tensor $p$ and the ground truth tensor $y$ is only computed within the precomputed lung segmentation as

$$L(p, y) = 1 - p.y$$

$$L(p, y) = 1 - \frac{p \cdot y + \epsilon}{p \cdot p + t \cdot t - p \cdot t + \epsilon}$$

Where $\epsilon = 1$ is the smoothing factor and $\cdot$ represents the tensor inner product operator. The loss function is optimized using Adabound (27) with an initial learning rate of $0.001$. The final model used for testing is selected using a validation set with 10 % of the training set patients that is randomly sampled.

The training and testing pipeline were implemented using Pytorch.